\documentclass[amsmath,amssymb,superscriptaddress,prl,twocolumn]{revtex4}

\usepackage{graphicx}
\usepackage{hyperref}
\usepackage{floatrow}

\hypersetup{
	bookmarks=false,
	pdfstartview={FitH},
	colorlinks=true,
	linkcolor=blue,
	citecolor=blue,
	urlcolor=blue}

\newcommand{\TaSe}{1\textit{T}-TaSe$_{2}$} 
\newcommand{\TaS}{1\textit{T}-TaS$_{2}$} 
\newcommand{\PLDsqrtthir}{$\sqrt{13} a_{0}\times \sqrt{13} a_{0}$} 
\newcommand{\Ag}{$A_{\mathrm{1g}}$} 
\newcommand{\Tcdw}{$T_{\mathrm{CDW}}$} 
\newcommand{\delMott}{$\Delta_{\mathrm{Mott}}$} 
\newcommand{\delCDW}{$\Delta_{\mathrm{CDW}}$} 
\newcommand{\qcdw}{$\textbf{q}_{\mathrm{CDW}}$} 
\newcommand{\mJcm}{mJ cm$^{-2}$} 
\newcommand{\cm}{cm$^{-1}$} 
\newcommand{\hv}{$h\nu$} 
\newcommand{\Ef}{$E_{\mathrm{F}}$} 
\newcommand{\kpara}{$k_{||}$} 
\newcommand{\overbar}[1]{\mkern 1.5mu\overline{\mkern-1.5mu#1\mkern-1.5mu}\mkern 1.5mu}
\newcommand{\GammaBar}{$\overbar{\mathrm{\Gamma}}$} 
\newcommand{\MBar}{$\overbar{\mathrm{M}}$} 
\newcommand{\KBar}{$\overbar{\mathrm{K}}$} 

\begin{document}

\title{Coherent phonons and the interplay between charge density wave and Mott phases in \TaSe}

\author{C.~J.~Sayers}
\email[Corresponding author: ]{charles.sayers@polimi.it}
\affiliation{Dipartimento di Fisica, Politecnico di Milano, 20133 Milano, Italy}
\affiliation{Centre for Nanoscience and Nanotechnology, Department of Physics, 
	University of Bath, Bath, BA2 7AY, UK}

\author{H.~Hedayat}
\affiliation{Dipartimento di Fisica, Politecnico di Milano, 20133 Milano, Italy}
\affiliation{IFN-CNR, Dipartimento di Fisica, Politecnico di Milano, 20133 Milano, Italy}

\author{A.~Ceraso}
\affiliation{Dipartimento di Fisica, Politecnico di Milano, 20133 Milano, Italy}

\author{F.~Museur}
\affiliation{Universit\'e de Lyon, ENS de Lyon, Universit\'e Claude Bernard,CNRS, Laboratoire de Physique, F-69342 Lyon, France}

\author{M.~Cattelan}
\affiliation{School of Chemistry, University of Bristol, Cantocks Close, Bristol BS8 1TS, UK}

\author{L.~S.~Hart}
\affiliation{Centre for Nanoscience and Nanotechnology, Department of Physics, 
	University of Bath, Bath, BA2 7AY, UK}

\author{L.~S.~Farrar}
\affiliation{Centre for Nanoscience and Nanotechnology, Department of Physics, 
	University of Bath, Bath, BA2 7AY, UK}

\author{S.~Dal Conte}
\affiliation{Dipartimento di Fisica, Politecnico di Milano, 20133 Milano, Italy}

\author{G.~Cerullo}
\affiliation{Dipartimento di Fisica, Politecnico di Milano, 20133 Milano, Italy}

\author{C.~Dallera}
\affiliation{Dipartimento di Fisica, Politecnico di Milano, 20133 Milano, Italy}

\author{E.~Da Como}
\affiliation{Centre for Nanoscience and Nanotechnology, Department of Physics, 
	University of Bath, Bath, BA2 7AY, UK}

\author{E.~Carpene}
\affiliation{IFN-CNR, Dipartimento di Fisica, Politecnico di Milano, 20133 Milano, Italy}

\begin{abstract}

\TaSe~is host to coexisting strongly-correlated phases including charge density waves (CDWs) and an unusual Mott transition at low temperature.  Here, we investigate coherent phonon oscillations in \TaSe~using a combination of time- and angle-resolved photoemission spectroscopy (TR-ARPES) and time-resolved reflectivity (TRR). Perturbation by a femtosecond laser pulse triggers a modulation of the valence band binding energy at the \GammaBar-point, related to the Mott gap, that is consistent with the in-plane CDW amplitude mode frequency. By contrast, TRR measurements show a modulation of the differential reflectivity comprised of multiple frequencies belonging to the distorted CDW lattice modes. Comparison of the temperature dependence of coherent and spontaneous phonons across the CDW transition shows that the amplitude mode intensity is more easily suppressed during perturbation of the CDW state by the optical excitation compared to other modes. Our results clearly identify the relationship of the in-plane CDW amplitude mode with the Mott phase in \TaSe~and highlight the importance of lattice degrees of freedom.
\end{abstract}

\maketitle

\section{Introduction}
Understanding the delicate interplay between co-operating or competing phases of matter in quantum materials is an ongoing goal of fundamental research \cite{Li2016,daSilvaNeto2014,Kogar2017}. Driving these systems out of equilibrium using an intense femtosecond laser pulse offers the possibility to transiently suppress forms of electronic and lattice order and monitor their recovery in real time \cite{Giannetti2016}. The characteristic dynamics of these processes allows a classification of materials in the time-domain \cite{Hellmann2012}, and has been used to disentangle the underlying mechanisms of complex phases found in cuprate superconductors \cite{Boschini2018} and exciton-lattice driven CDW systems \cite{Hedayat2019}, or to unlock normally hidden states of matter \cite{Sun2018}. 

An ideal platform to investigate these phenomena are the trigonal (1\textit{T}) tantalum-based transition metal dichalcogenides (TMDs), MX$_{2}$ (M = Ta, X = S/Se) which are host to a whole range of strongly-correlated behaviour including charge density waves (CDWs) \cite{Wilson1975}, Mott physics \cite{Perfetti2003}, possible quantum spin liquid states \cite{Law2017}, and superconductivity \cite{Sipos2008,Liu2016}. 

Tantalum disulphide (\TaS) has been studied extensively using time-resolved techniques \cite{Perfetti2006,Perfetti2008,Eichberger2010,Peterson2011,Stojchevska2014,Sohrt2014,Ligges2018}, motivated mostly by its rich phase diagram. It exhibits multiple CDW transitions including an incommensurate (550 K), nearly-commensurate (350 K), and commensurate (180 K) phase \cite{Wilson1975} that occurs concomitantly with a metal-insulator transition, commonly associated with a Mott phase \cite{Fazekas1980}. However, an alternative explanation has recently been proposed based on the three-dimensional stacking order of the CDW and hybridization of atomic orbitals \cite{Ritschel2018,Lee2019,Stahl2020}. Thus, important questions surrounding the nature of the metal-insulator transition and its relationship with the CDW remain.

Tantalum diselenide (\TaSe)~has received comparatively far less attention, although it was recently suggested to be the more suitable compound to investigate the relationship between the CDW and Mott phases because of the well-separated transition temperatures, larger electronic gap, and reduced complexity due to the absence of the nearly-commensurate phase (NCCDW) \cite{Sohrt2014}. 

\TaSe~undergoes a first-order transition from an incommensurate (ICCDW) to commensurate (CCDW) charge density wave at \Tcdw~= 473 K \cite{DiSalvo1974}. It is accompanied by an in-plane \PLDsqrtthir~periodic lattice distortion (PLD) which is rotated by $\sim$~13$^{\circ}$ with respect to the original unit cell, and forms a 13-atom superlattice comprised of Ta clusters in the so-called ``star-of-David" configuration \cite{Wilson1975}.

In addition to the well known CDW, a Mott transition occurs at $\sim$~260 K evidenced by the opening of a gap, \delMott~$\approx$~250 meV below the Fermi level, \Ef~observed by ARPES \cite{Perfetti2003} and STM \cite{Colonna2005}. The CDW has been suggested to be a precursor to the Mott phase, since it modifies the band structure resulting in a narrow half-filled band at \Ef. As the temperature is reduced, the increasing CDW amplitude causes a narrowing of the band width ($W$), related to the in-plane electron hopping between the adjacent star-of-David clusters in the CDW lattice \cite{Colonna2006,Chen2020}. Electrons will become localized when the band width ($W$) decreases below a critical value and the Mott criterion, $U/W \geq 1$ is reached, where $U$ is the on-site electron-electron Coulomb repulsion. The result of this localization is a transition to an insulating state and formation of a gap, \delMott~\cite{Perfetti2003,Colonna2005}.

Time- and angle-resolved photoemission spectroscopy (TR-ARPES) is a powerful tool which allows direct visualization of the electronic band structure following perturbation with a laser pulse. In strongly-correlated systems, it can be used to monitor the collapse and recovery of electronic order in real-time by probing energy gaps, $\Delta$ related to the order parameter \cite{Hellmann2012,Boschini2018,Hedayat2019}. Recent TR-ARPES studies of \TaSe~using high-harmonics (\hv~$\approx$ 22 eV) have focused predominantly on gap suppression dynamics on short timescales \cite{Sohrt2014} or the room temperature (CDW) phase only \cite{Shi2019}, and thus the Mott phase dynamics remain relatively unexplored.

Here, we report on electron and phonon dynamics of \TaSe~at low temperature (77 K) where the CDW and Mott phases co-exist. Using TR-ARPES with \hv~= 6 eV photon energy, we track the temporal evolution of the valence band binding energy related to \delMott~at the \GammaBar-point over several picoseconds. We find that it exhibits strong, long-lasting oscillations with a single frequency related to the in-plane CDW amplitude mode ($\sim$~2.2 THz at 77 K). Using complementary time-resolved reflectivity (TRR) measurements, we instead find multiple phonon frequencies related to the PLD ($\sim$~1.8, 2.2 and 2.9 THz). Therefore, aided by the momentum-selectivity of TR-ARPES, our results reveal that the gap, \delMott~is linked preferentially to the amplitude mode of the CDW. By investigating the temperature-dependence of coherent phonons, we find that the amplitude mode deviates significantly from first-order behaviour, whilst the other modes are robust up to \Tcdw, highlighting further the nature of this mode and its importance to electronic order in this material.

\begin{figure}[ht!]
	\begin{center}
		\includegraphics[width=1.0\linewidth,clip]{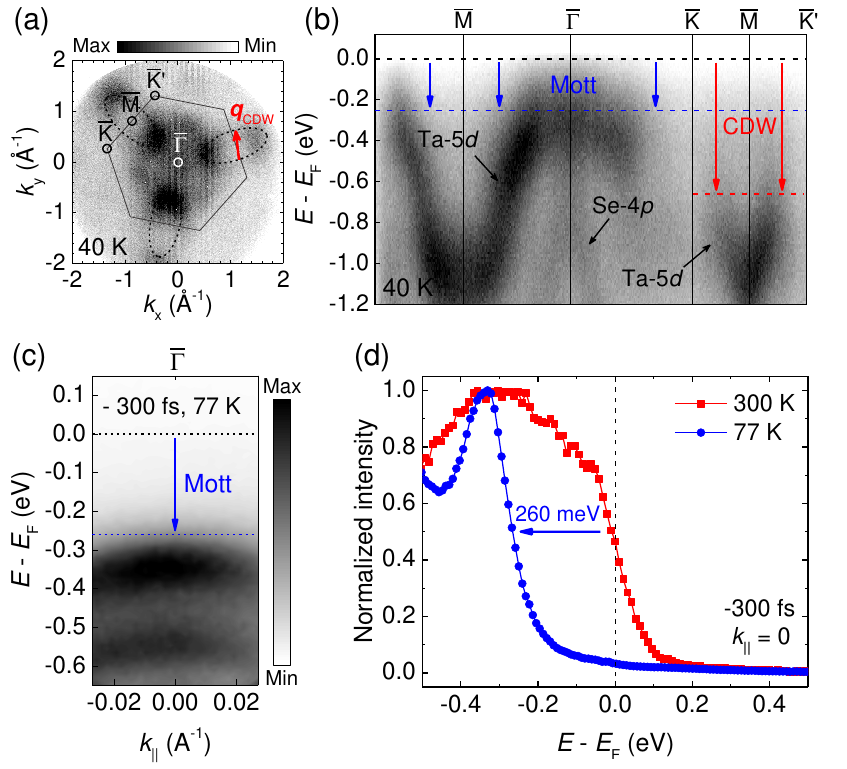}
	\end{center}
	\vspace*{-0.2in}
	\caption{\label{fig:Order}Electronic order in \TaSe. (a) Full-wavevector ARPES (\hv~=~21.2 eV) image of the electronic structure, $E - E_{\mathrm{F}}$~=~-0.5 eV at 40 K with projected high-symmetry points of the hexagonal Brillouin zone (BZ) labelled. The red arrow is the expected CDW vector, \qcdw. (b) Band dispersions through the BZ. The vertical arrows highlight the lowering of occupied states due to the CDW (red) and Mott (blue) transitions. (c) TR-ARPES (\hv~=~6 eV) map at 77 K. (d) Comparison of EDCs extracted from the $\overbar{\Gamma}$-point (\kpara = 0) at 300 K and 77 K. The arrow shows the band edge shift as a result of the Mott transition.}
\end{figure}

\section{Methods}
TR-ARPES experiments were performed using visible ($\sim$1.8 eV, 30 fs) pump and deep-ultraviolet ($\sim$6.0 eV, 80 fs) probe pulses generated by a series of nonlinear optical processes from the output of an Yb-based laser (Pharos, Light Conversion) operating at 80 kHz repetition rate, as described in Ref. \cite{Boschini2014}. The overall time- and energy-resolution of this configuration were approximately 80 fs and 45 meV, respectively. TRR was performed using a setup based on a Ti:sapphire laser (Coherent Libra) which drives two non-collinear parametric amplifiers (NOPAs) serving as pump and probe beams \cite{Manzoni2006}. The amplified pulses are characterized by a broad spectrum of (1.8 - 2.4) eV and compressed to $\leq$~20 fs duration using chirped mirrors. Steady-state ARPES measurements were performed at the Bristol NanoESCA facility using He-I$\alpha$ radiation (\hv~=~21.2 eV), with $\sim$50 meV energy resolution at 40 K. Further experimental details relating to crystal growth methods, electrical transport measurements, and Raman spectroscopy are provided in the Supplemental Material (Ref.~\cite{Supp}).

\section{Results and discussion}
First we discuss the electronic structure of \TaSe~and the effects of CDW and Mott transitions. Fig.~\ref{fig:Order}(a) shows a full-wavevector ARPES image at $E - E_{\mathrm{F}}$~=~-0.5 eV where all the main features are visible. Centred around the \MBar-points on the BZ boundary are the elliptical Ta-5\textit{d} electron pockets, and at the BZ centre (\GammaBar-point) is the Se-4\textit{p} pocket, in agreement with previous ARPES measurements \cite{Clerc2004,Bovet2004}. The CDW involves finite portions of the Ta-5\textit{d} electron pockets which are linked by the wavevector, \qcdw~in the \KBar-\MBar-\KBar~direction where the Fermi surface could be prone to nesting \cite{Wilson1975,Sohrt2014}, although the importance of such electronic instabilities is still debated \cite{Aebi2001,Bovet2004}. Indeed, there is a clear loss of intensity on the parallel arms of these pockets in Fig.~\ref{fig:Order}(a), and dispersion along the \KBar-\MBar-\KBar~direction in Fig.~\ref{fig:Order}(b) shows the band edge is found significantly below \Ef~as a result of the CDW gap, \delCDW. Dispersion along the \MBar-\GammaBar~direction in Fig.~\ref{fig:Order}(b) shows a valence band comprised of broad Ta-5\textit{d} states, and steeply dispersing Se-4\textit{p} states, which are known to extend to \Ef~at room temperature \cite{Perfetti2003}. At 40 K, all bands near \Ef~have been lowered by $\sim$250 meV due to a gap, \delMott~which extends across all $k$-space, and thus the entire Fermi surface is removed \cite{Supp}.

\begin{figure}[t!]
	\begin{center}
		\includegraphics[width=1.0\linewidth,clip]{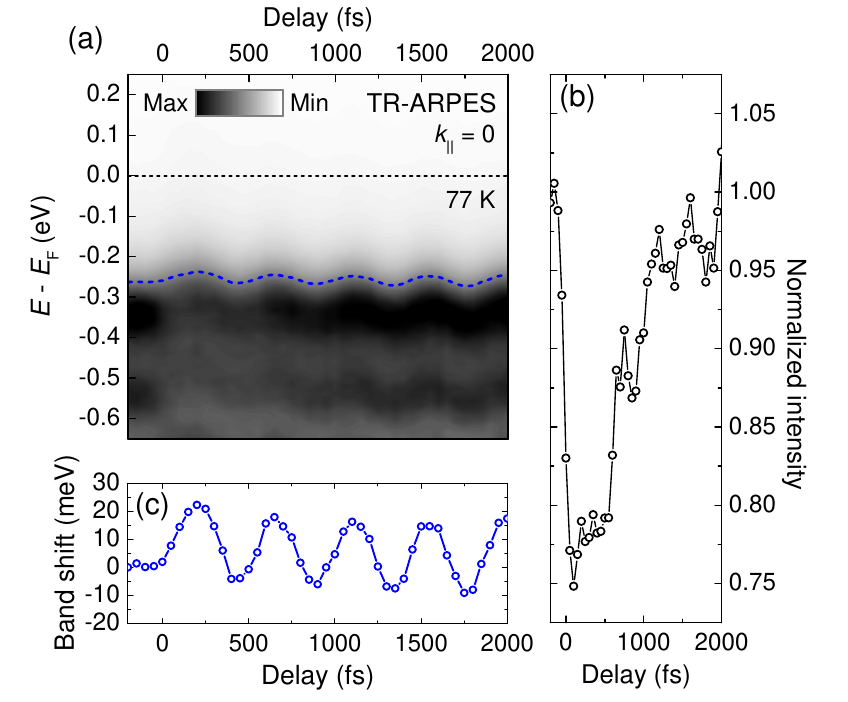}
	\end{center}
	\vspace*{-0.2in}
	\caption{\label{fig:Response}Valence band dynamics in the CDW-Mott phase. (a) TR-ARPES spectra at the \GammaBar-point (77 K) using 1.10 \mJcm~pump fluence. (b) Normalized valence band intensity, extracted from the maximum near $E - E_{\mathrm{F}}$ $\approx$ -0.35 eV. (c) Valence band shift extracted from a constant intensity contour in panel (a), indicated by the blue dashed line.}
\end{figure}

Shown in Fig.~\ref{fig:Order}(c) are TR-ARPES spectra of \TaSe~at 77 K near the \GammaBar-point before pump arrival (- 300 fs). The dispersion is approximately along the \MBar-\GammaBar-\MBar~direction based on low energy electron diffraction (LEED)~\cite{Supp}. Fig.~\ref{fig:Order}(c) clearly shows the band edge is situated below \Ef~due to the opening of a gap. A comparison of the energy distribution curves (EDCs) in Fig.~\ref{fig:Order}(d) extracted at \GammaBar~(\kpara~= 0) shows a lowering of the band edge by $\sim$~260 meV between 300 K and 77 K, which is in very close agreement with previous reports of the Mott transition in \TaSe~\cite{Perfetti2003} and the ARPES measurements in Fig.~\ref{fig:Order}(b).

We note that such a substantial modification of the Fermi surface, as seen in Fig~\ref{fig:Order}(c) and (d), typically indicates the development of an insulating state, which would be expected to be observed by electrical transport \cite{Knowles2020}. However, our resistance measurements show metallic behaviour to 4 K \cite{Supp}. These contrasting results between surface probe (ARPES) and bulk probe (transport) have previously provided support for the hypothesis that the Mott transition in \TaSe~is only a surface effect \cite{Perfetti2003}. However, we emphasize that for TR-ARPES at \hv~= 6 eV, the inelastic mean free path for electrons is expected to be of the order $\sim$ 10 nm \cite{Seah1979} and hence the photoemission signal may represent several layers of \TaSe~given the $c$-axis length of 0.63 nm \cite{DiSalvo1974}. Thus, the Mott phase may extend over more than the surface layers.

We now focus on the valence band dynamics in the coexisting CDW-Mott phase measured by TR-ARPES. Fig.~\ref{fig:Response}(a) shows the valence band at the \GammaBar-point after perturbation by the pump pulse. At $t$ = 0, the optical excitation results in an instantaneous loss of valence band intensity as highlighted in Fig.~\ref{fig:Response}(b), and recovery occurs within $\sim$~2 ps which is similar to the reported dynamics of \TaS~in the Mott phase \cite{Perfetti2006}. As is clearly evident by the persisting gap, \delMott~in Fig.~\ref{fig:Response}(a), we do not observe a collapse of the Mott phase and we also note that 1.10~\mJcm~pump fluence is not sufficient to melt the CDW \cite{Ji2020}. Hence, we confirm that the TR-ARPES experiments were performed in the coexisting CDW-Mott phase.

\begin{figure*}[ht!]
	\begin{center}
		\includegraphics[width=1.0\linewidth,clip]{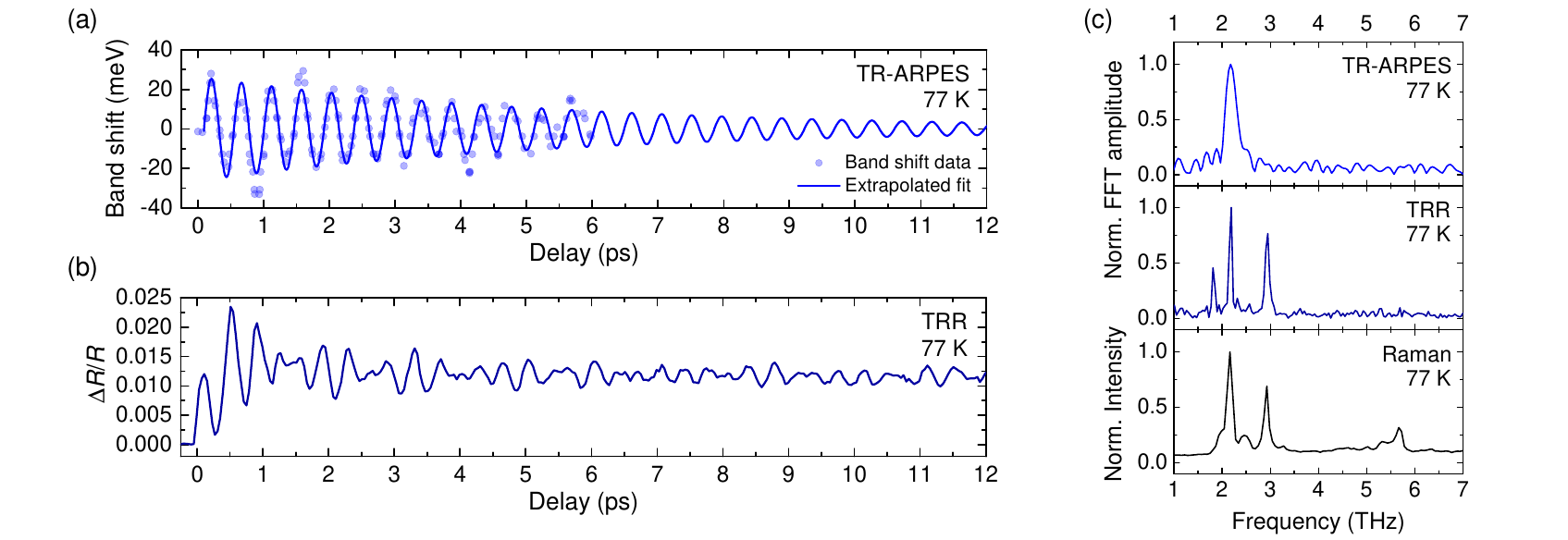}
	\end{center}
	\vspace*{-0.2in}
	\caption{\label{fig:Oscillations} Coherent phonon oscillations in the CDW-Mott phase. (a) Oscillatory component of the valence band shift measured by TR-ARPES (1.16~\mJcm). (b) Transient reflectivity signal measured by TRR (0.11~\mJcm), where $\Delta R / R$ is the absolute value of the differential reflectivity. The selected data is for 1.84 eV probe photon energy. (c) Normalized fast Fourier transform (FFT) amplitude of the TR-ARPES and TRR oscillatory components, together with a Raman spectrum for comparison.}
\end{figure*}

Fig.~\ref{fig:Response}(c) shows the temporal evolution of the valence band edge position. Most noticeably, the pump triggers strong coherent oscillations which are weakly damped. TR-ARPES data in Fig.~\ref{fig:Oscillations}(a) shows a maximum initial oscillation of approximately $\pm$ 20 meV around the equilibrium position with a large amplitude that persists at 6 ps. Fitting the data with a damped periodic function $E(t) = A \exp \left( -t / \tau_d \right) \sin \left( 2 \pi \omega t + \phi \right)$ yields a frequency, $\omega \approx$ (2.19 $\pm$ 0.01) THz, which closely matches the intense 72.4 \cm~(2.17 THz) \Ag~mode measured by Raman spectroscopy at 77 K \cite{Supp}. The damping time was found to be $\tau_d  \approx$ (6.3 $\pm$ 1.0) ps. Such long-lived oscillations have also been observed in the Mott phase of \TaS~\cite{Perfetti2006,Perfetti2008} and were assigned to the CDW amplitude mode, related to the in-plane breathing mode of the stars-of-David \cite{Demsar2002}. The result presented here shows a direct modulation of the binding energy of the valence band edge, related to \delMott, by the CDW amplitude mode in \TaSe. This is consistent with the CDW precursor scenario whereby the CDW amplitude, related to the magnitude of the in-plane PLD, controls the $U/W$ criterion by the degree of electron hopping between the adjacent stars-of-David \cite{Sohrt2014,Fazekas1980,Colonna2005,Colonna2006,Chen2020}. By triggering coherent oscillations of the CDW amplitude (breathing) mode, we induce a modulation in the $U/W$ ratio which manifests in the magnitude of \delMott.

To investigate the origin of the coherent phonon oscillations further, we now compare our TR-ARPES results with TRR measurements. Fig.~\ref{fig:Oscillations}(b) shows the temporal evolution of the differential reflectivity, $\Delta R / R$ of \TaSe~at 77 K. The $\Delta R / R$ signal is dominated by strong oscillations that are weakly damped and last up to 20 ps. Interestingly, the oscillations in TRR are comprised of multiple frequencies, in stark contrast to the single-frequency valence band modulation observed by TR-ARPES. This is confirmed by the fast Fourier transform (FFT) of the oscillatory components shown in Fig.~\ref{fig:Oscillations}(c). The FFT of the valence band dynamics shows a single frequency at $\sim$~2.2 THz, whereas FFT of the TRR signal shows multiple frequencies with greatest amplitude at $\sim$~1.8, 2.2 and 2.9 THz, and closely resembles the Raman spectrum presented in  Fig.~\ref{fig:Oscillations}(c) and reported previously \cite{Smith1976,Tsang1977,Uchida1981,Sugai1981}. We note that the $\sim$~1.8 THz mode cannot be seen in the Raman data as it falls below the cut-off of the spectrometer laser filter. In addition, we note that the 2.2 THz mode appears broader in the FFT of the TR-ARPES data because of the shorter sampling interval of the oscillations. Since both TR-ARPES and TRR experiments utilize comparable pump photon energies and pulse durations, it is conceivable that multiple modes are triggered in both cases and relate to the Raman-active $\Gamma$-point phonons of the PLD. The energy-momentum selectivity of TR-ARPES directly probes the local electronic structure of the valence band at \GammaBar~and the interactions there. Hence, the observed modulation of the valence band binding energy (\delMott) with a single frequency belonging to the $\sim$~2.2 THz CDW amplitude mode, shows that the Mott phase is preferentially linked to that particular mode.

\begin{figure}[ht!]
	\begin{center}
		\includegraphics[width=1.0\linewidth,clip]{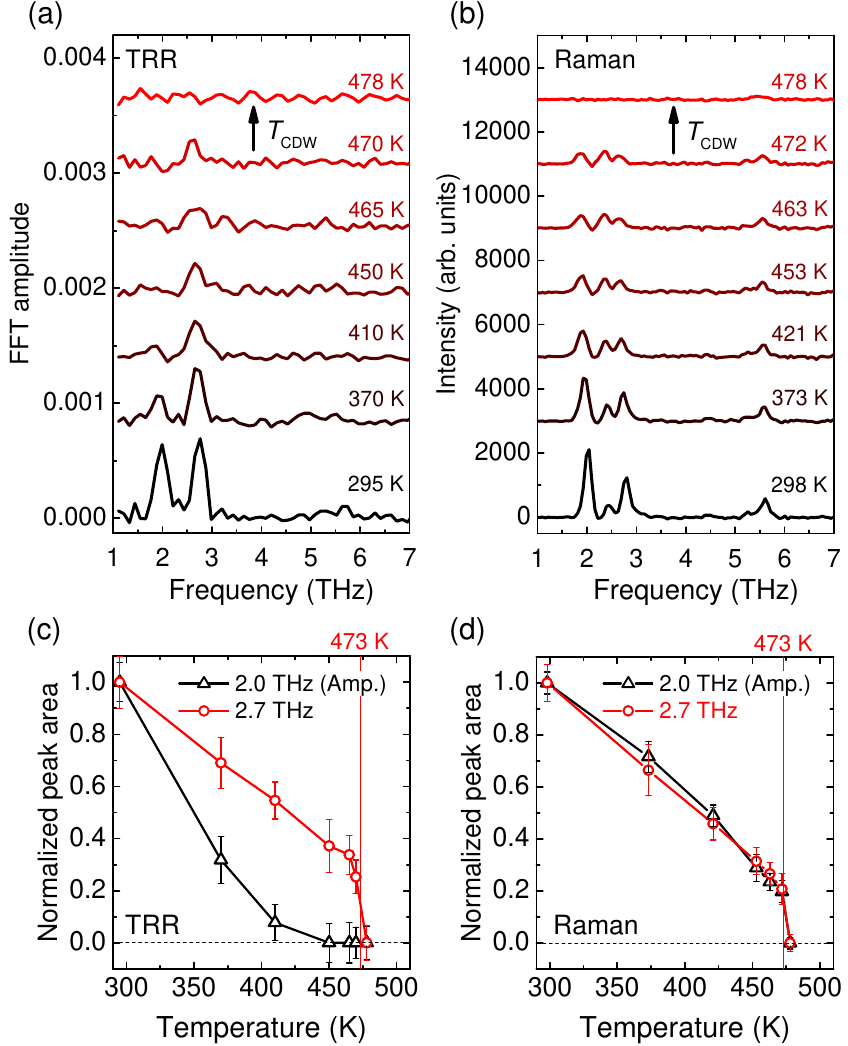}
	\end{center}
	\vspace*{-0.2in}
	\caption{\label{fig:TempDep}Temperature dependence of coherent and spontaneous phonons in the CDW phase. (a) Fast Fourier transform (FFT) of the transient reflectivity, $\Delta R / R$ signal measured by TRR (0.11~\mJcm) at sample temperatures as indicated. The selected data is for 655 nm probe wavelength. (b) Raman spectra measured over a similar temperature range as panel (a) for comparison, after subtraction of a fit to the incoherent background signal above \Tcdw~(478 K). All traces are offset for clarity. Panels (c) and (d) show the temperature dependence of the integrated peak area for the 2.0 amplitude (Amp.) and 2.7 THz modes in the TRR-FFT and Raman spectra respectively.}
\end{figure}

Having established the coherent phonon oscillations of the CDW lattice and the single mode which is linked to the Mott phase, we finally focus on the temperature dependence of these modes, and their behaviour across the CDW transition at \Tcdw. For this, we compare the response of the coherent phonons to optical excitation by TRR, and the spontaneous phonons of the PLD in quasi-equilibrium by Raman spectroscopy. Shown in Fig.~\ref{fig:TempDep}(a) is a FFT analysis of the $\Delta R / R$ signal for various sample temperatures in the range (295 - 478) K which are compared to Raman spectra in Fig.~\ref{fig:TempDep}(b). Similar to the TRR data at 77 K, multiple frequency components are found at 295 K as shown in Fig.~\ref{fig:TempDep}(a). The peaks in FFT amplitude belong to the two highest intensity modes determined previously [see Fig.~\ref{fig:Oscillations}(c)], although they are found at slightly lower frequencies of $\sim$~2.0 and 2.7 THz because of the higher sample temperature \cite{Supp}. Fig.~\ref{fig:TempDep}(a) shows that as the temperature increases in the range (295 - 410) K, the intensity of the $\sim$~2.0 THz amplitude mode decreases sharply until it becomes absent for $T$ $\geq$ 450 K using 0.11~\mJcm~fluence. Instead, the $\sim$~2.7 THz mode remains present until heating above the first-order ICCDW-CCDW phase transition at \Tcdw~= 473 K. By comparison, the Raman data in Fig.~\ref{fig:TempDep}(b) shows that all modes remain clearly visible until there is a sudden change in the spectra when heating above \Tcdw. Specifically, we find that all modes merge into a broad background (see Ref.~\cite{Supp}), similar to previous reports \cite{Tsang1977,Smith1976,Uchida1981,Sugai1981}. The stark difference in the temperature dependence of the mode intensities measured by the two experimental techniques is highlighted by comparing Figs.~\ref{fig:TempDep}(c) and (d) which show the integrated peak areas in TRR and Raman spectroscopy, respectively. The expected first-order nature of the CDW transition is clear in Fig.~\ref{fig:TempDep}(d) whereby there is a steep onset at \Tcdw~followed by linear temperature dependence, and both the $\sim$~2.0 and 2.7 THz modes exhibit identical behaviour. Instead, Fig.~\ref{fig:TempDep}(c) shows a dramatic suppression of the $\sim$~2.0 THz amplitude mode intensity and a deviation from first-order behaviour in TRR, suggestive of a transient photoinduced melting of the CDW amplitude. The $\sim$~2.7 THz mode however, which is a phonon of the PLD \cite{Tsang1977}, appears to remain robust up to \Tcdw. A complete loss of intensity of the CDW amplitude mode suggests that it has become strongly damped, whereby its lifetime is less than the period of oscillation ($\approx$ 0.5 ps). Such increased damping could be due to a reduced commensurability between the CDW and the underlying lattice which results in a faster dephasing of the oscillations \cite{Perfetti2008}, providing evidence for a suppression of the commensurate state by the optical excitation before the ICCDW-CCDW transition at \Tcdw.

\section{Conclusion}
In summary, an investigation of electron and phonon dynamics in the coexisting CDW-Mott phase of \TaSe~using complementary TR-ARPES and TRR techniques clearly shows that the Mott phase is preferentially linked to the in-plane CDW amplitude, since it controls the degree of electron localization between adjacent star-of-David configurations. Our results highlight the role of the CDW and lattice degrees of freedom in stabilizing the Mott phase of \TaSe~and further the understanding of the interplay between these coexisting phases.

\section{Acknowledgements}
The authors acknowledge funding and support from the EPSRC Centre for Doctoral Training in Condensed Matter Physics (CDT-CMP) Grant No.~EP/L015544/1 and the Italian PRIN Project No.~2017BZPKSZ. We acknowledge the School of Chemistry at the University of Bristol is for access to the Bristol NanoESCA Facility. Finally, we thank Daniel Wolverson for useful discussions.

\bibliographystyle{unsrtnat}

\clearpage
\onecolumngrid
\begin{center}
	\textbf{\large Supplemental material for: ``Coherent phonons and the interplay between charge density wave and Mott phases in \TaSe"}
\end{center}

\setcounter{equation}{0}
\setcounter{figure}{0}
\setcounter{table}{0}
\setcounter{section}{0}
\setcounter{page}{1}
\makeatletter
\renewcommand{\thefigure}{S\arabic{figure}}

\section{Crystal growth}\label{Sec:Crystal growth}
\begin{figure}[h]
	\floatbox[{\capbeside\thisfloatsetup{capbesideposition={right,center},capbesidewidth=8cm}}]{figure}[\FBwidth]
	{\includegraphics[width=4cm]{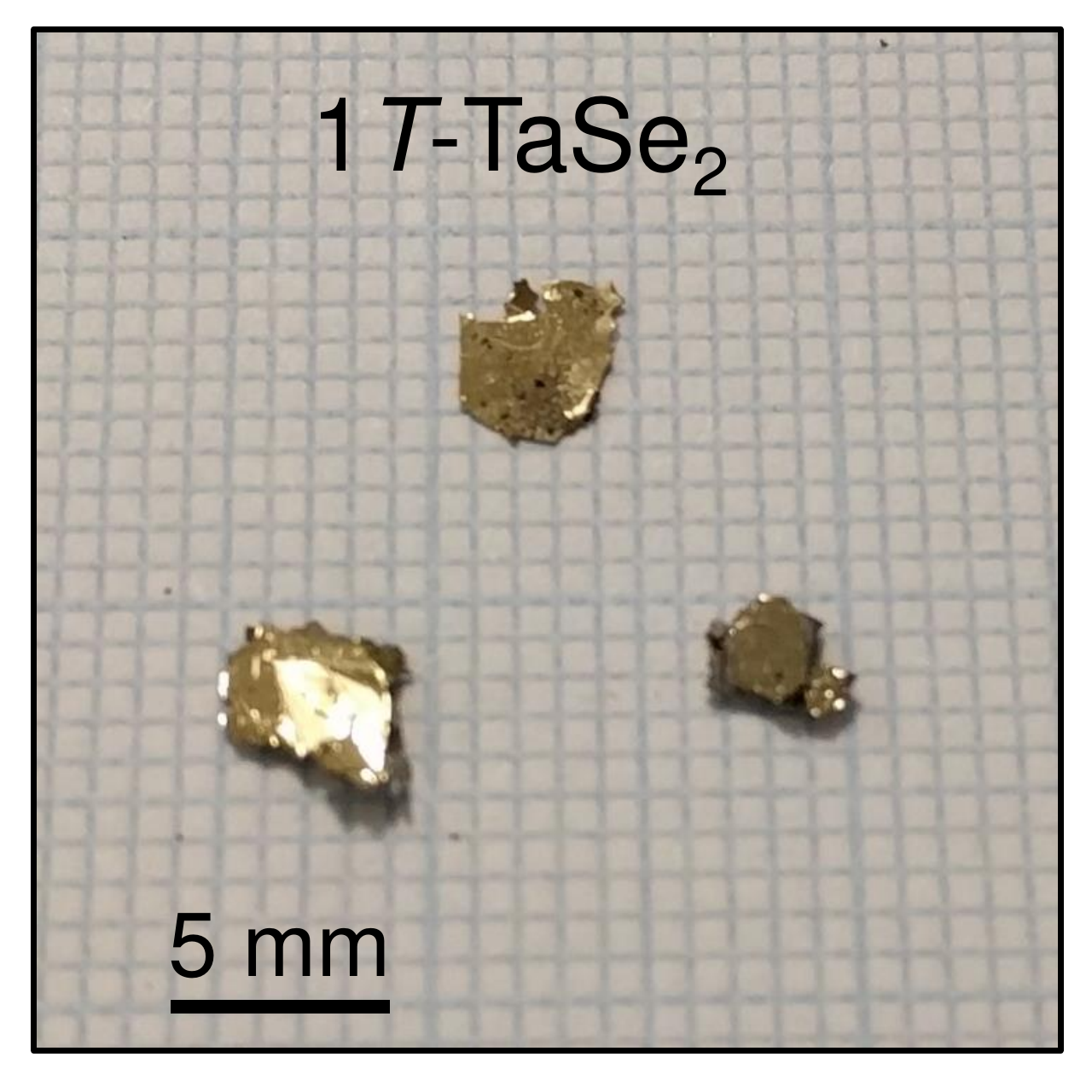}}
	{\caption{\label{fig:Growth} Single crystals of \TaSe. Samples were grown by the chemical vapour transport (CVT) method. The image shows typical crystals which were selected for TR-ARPES measurements.}}
\end{figure}

Single crystals of \TaSe~were grown using the chemical vapour transport (CVT) technique inside an evacuated quartz ampoule. The starting materials consisted of high-purity tantalum (99.9\%) and selenium (99.9\%) powders, together with pellets of anhydrous iodine (99.9\%) which acts as the transport agent. Crystals were grown in a slight excess of selenium ($\sim$ 3 mg cm$^{-3}$) to encourage ideal stoichiometry \cite{suppDiSalvo1974}. A growth temperature of $T_{\mathrm{g}}$ = 960$^{\circ}$C~was maintained for $\sim$ 21 days before the ampoule was quenched rapidly in water from high temperature in order to retain the metastable 1\textit{T} phase, similar to the methods of Ref.~\cite{suppDiSalvo1974}. The resulting crystals were up to (4 $\times$ 4) mm across with regular hexagonal edges and had a metallic gold colour characteristic of \TaSe, as shown in Fig.~\ref{fig:Growth}.

\section{Electronic transport}\label{Sec:Electronic transport}
\begin{figure}[H]
	\floatbox[{\capbeside\thisfloatsetup{capbesideposition={right,center},capbesidewidth=5cm}}]{figure}[\FBwidth]
	{\includegraphics[width=10cm]{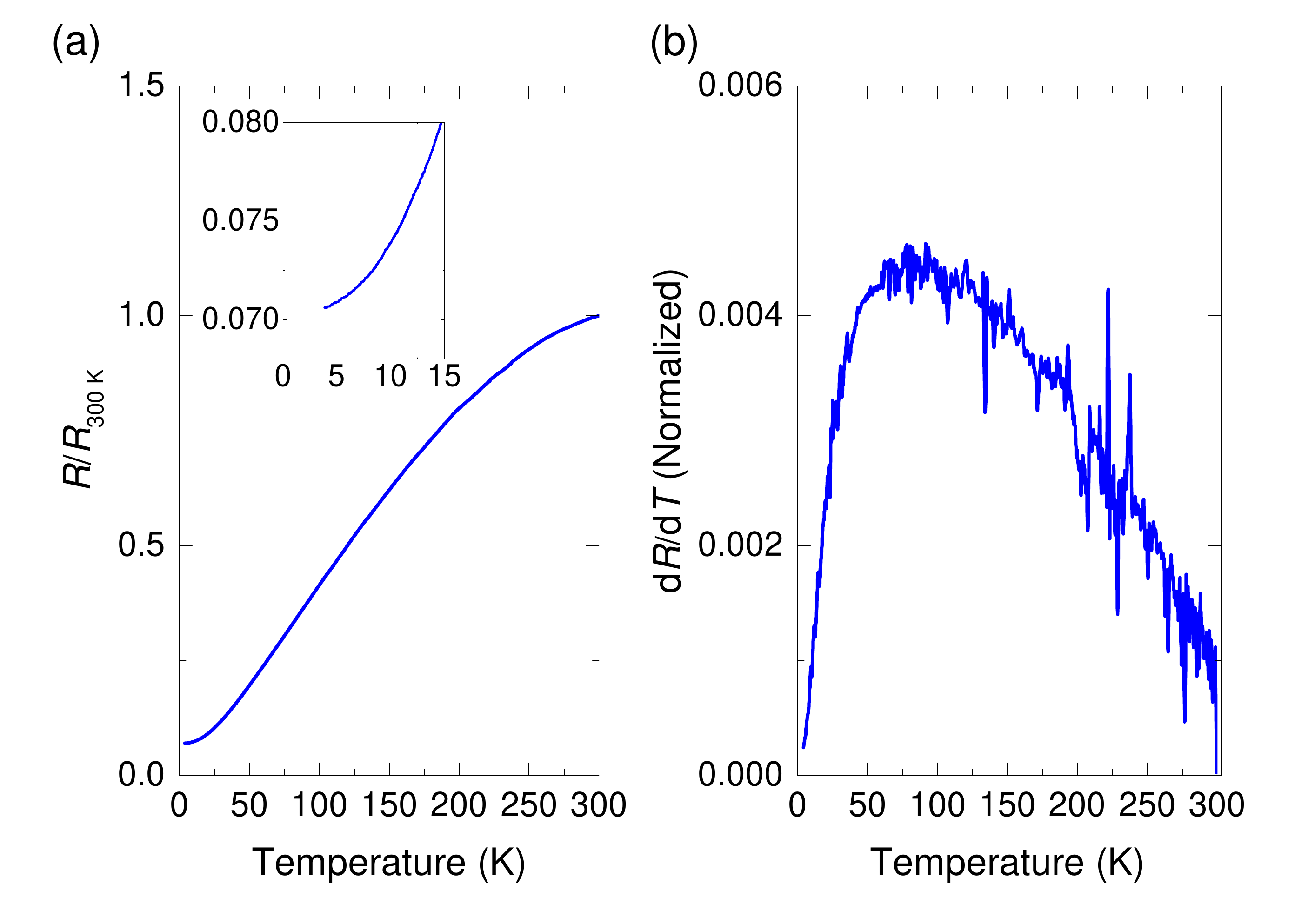}}
	{\caption{\label{fig:Transport}  Electronic transport. (a) Normalized resistance, $R$/$R_{\mathrm{300K}}$ as a function of temperature. The inset shows the low temperature range ($<$ 15 K). (b) First derivative of resistance, d$R$/d$T$.}}
\end{figure}

Fig.~\ref{fig:Transport}(a) shows electrical resistance as a function of temperature for \TaSe~crystals selected from the same batch as those used for TR-ARPES measurements. The shape of the curve is typical for this material across the measured temperature range of (300 - 4) K with overall metallic behaviour \cite{suppDiSalvo1974}.

By analysing the first derivative of resistance, d$R$/d$T$ in Fig.~\ref{fig:Transport}(b), there are no obvious signatures of phase transitions. This is in stark contrast to the clear metal-insulator type (Mott) transition observed by ARPES, whereby a significant gap of $\sim$ 250 meV below \Ef~develops across all $k$-space upon cooling from 300 K to 77 K. Following this, one would expect a large increase in resistance near the Mott transition temperature, $T_\mathrm{Mott} \approx$ 260 K~\cite{suppPerfetti2003}, and possibly a crossover to insulating behaviour, which is clearly not observed in Fig.~\ref{fig:Transport}. These contrasting results are discussed in the main text in relation to the hypothesis of Perfetti \textit{et al.}~\cite{suppPerfetti2003} that the Mott transition in \TaSe~is only a surface effect.

The residual resistance ratio (RRR), $R_{\mathrm{300K}}$/$R_{\mathrm{4K}}$ allows an estimation of the crystal quality. We find a range of RRR $\approx$ (14 - 16) within the batch, which indicates reasonable quality compared to previous reports \cite{suppDiSalvo1974}. The inset of Fig.~\ref{fig:Transport}(a) shows that there is no upturn in $R(T)$ at low temperatures which one would expect if there were significant defect scattering.

\section{Raman spectroscopy}\label{Sec:Raman spectroscopy}
\begin{figure}[h]
	\begin{center}
		\includegraphics[width=0.75\linewidth,clip]{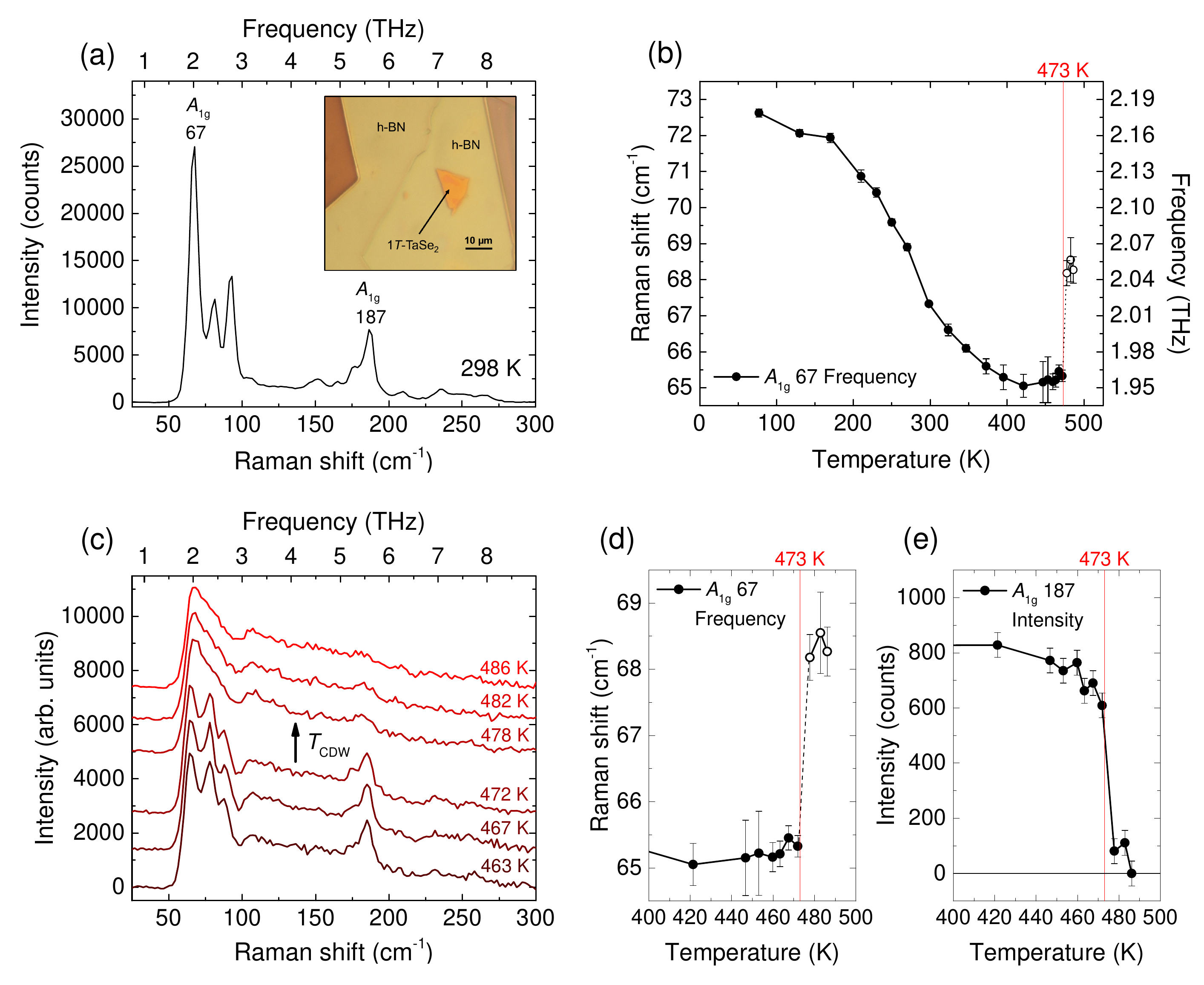}
	\end{center}
	\vspace*{-0.2in}
	\caption{\label{fig:Raman} Temperature dependence of phonon modes in \TaSe~measured by Raman spectroscopy. (a) Raman spectrum at room temperature (298 K). The inset shows a bulk-like \TaSe~flake encapsulated between sheets of h-BN which was used to measure the high-temperature range (298 - 486) K. (b) Temperature dependence of the \Ag~67 mode in the range (77 - 486) K. (c) Raman spectra near the expected CDW phase transition (\Tcdw~$\approx$ 473 K). (d) Frequency of the \Ag~67 mode near the expected phase transition. Error bars are the standard deviation from the Voigt fitting only. (e) Intensity of the \Ag~187 mode above the background level. The error bars represent the magnitude of the background noise in the spectra near \Ag~187.}
\end{figure}

In order to determine the CDW phase transition and to monitor the temperature dependence of phonon modes in \TaSe, Raman spectroscopy was performed in the range (77 - 486) K as shown in Fig.~\ref{fig:Raman}. 

The low temperature data (77 - 270 K) was measured on a bulk crystal mounted inside a flow cryostat under vacuum. Instead, the high temperature data (298 - 486 K) was acquired from a sample mounted on a heating plate in air. To avoid oxidation at high temperature in the latter case, the \TaSe~sample was encapsulated in hexagonal boron nitride (h-BN). A bulk-like \TaSe~flake was obtained by exfoliation from single crystals ($>$ 20 layers based on optical contrast) which was then encapsulated between two flakes of h-BN (2D-semiconductors.com) on a Si/SiO$_{2}$ substrate using a dry transfer technique \cite{suppCastellanos-Gomez2014} in an oxygen-free environment ($<$ 0.5 ppm) as shown in the inset of Fig.~\ref{fig:Raman}(a). Measurements were made using a Renishaw InVia Raman spectrometer operating in backscattering geometry. The laser excitation wavelength was 532 nm and the grating was 1800 l/mm. A long working-distance 50$\times$~objective lens (N.A. = 0.5) was used. The overall spectral resolution was better than 1 \cm. The laser power on the sample was (65 $\pm$ 1) $\mu$W to minimize heating.

Similar to other TMDs with trigonal (1\textit{T}) structure, \TaSe~has only two Raman-active modes in the normal phase (i.e. undistorted lattice in the absence of CDW) with $A_{\mathrm{g}}$ and $E_{\mathrm{g}}$ symmetry \cite{suppTsang1977}. In the CDW phase, the distorted lattice is triclinic and there are 114 optical modes in total, of which 57 are Raman-active with $A_{\mathrm{g}}$ symmetry \cite{suppSugai1981}. Fig.~\ref{fig:Raman}(a) shows a spectrum of \TaSe~in the CDW phase (298 K) with a particularly high signal-to-noise ratio where it is possible to identify 12 individual modes. For the purpose of the following analysis, two particular modes are labelled according to their peak centre at room temperature, namely \Ag~67 and \Ag~187.

In Fig.~\ref{fig:Raman}(b), the frequency of the \Ag~67 ($\sim$ 2 THz) mode is shown as a function of temperature. This mode is very strongly temperature-dependent and shifts $>$ 7 \cm~between 77 K and 470 K. By contrast, the frequency of the \Ag~187 mode is almost temperature independent. Shown in Fig.~\ref{fig:Raman}(c) are the Raman spectra near the expected CDW phase transition (\Tcdw~$\approx$ 473 K). Here, there is a sudden change in the spectrum upon heating between 472 and 478 K. A loss of definition of all the modes occurs and only a broad background remains below 100 \cm, which is similar to previous reports and indicates the sample has passed from the CCDW to ICCDW phase \cite{suppTsang1977}. We note that this behaviour is also similar to \TaS~when passing from the CCDW to the NCCDW \cite{suppSmith1976}. Such a sudden loss of CDW order across a narrow temperature range is characteristic of a first-order phase transition in these compounds \cite{suppDiSalvo1974,suppTsang1977,suppSugai1981}. By contrast, other 1\textit{T}-TMDs such as 1\textit{T}-TiSe$_{2}$ and 1\textit{T}-VSe$_{2}$ exhibit softer (second-order) development of the CDW phase whereby the phonons of the distorted lattice are only well-resolved far below the transition temperature ($\sim$~60 - 80 K $<$ \Tcdw) \cite{suppSmith1976,suppUchida1981,suppHedayat2019}. Fig.~\ref{fig:Raman}(d) and (e) show the temperature dependence of the \Ag~67 frequency and the \Ag~187 intensity respectively over the range (400 - 500) K. The sudden change near 473 K (red vertical line) confirms the expected phase transition temperature in our samples. In Fig.~\ref{fig:Raman}(d), the dashed line joining the data points above and below \Tcdw~indicates that the plotted frequency is the peak of the well-defined \Ag~mode below~\Tcdw, whereas it is the peak of the broad feature above \Tcdw, which can no longer be unambiguously identified as the same mode.

\section{LEED}\label{Sec:LEED}
\begin{figure}[H]
	\floatbox[{\capbeside\thisfloatsetup{capbesideposition={right,center},capbesidewidth=8cm}}]{figure}[\FBwidth]
	{\includegraphics[width=6cm]{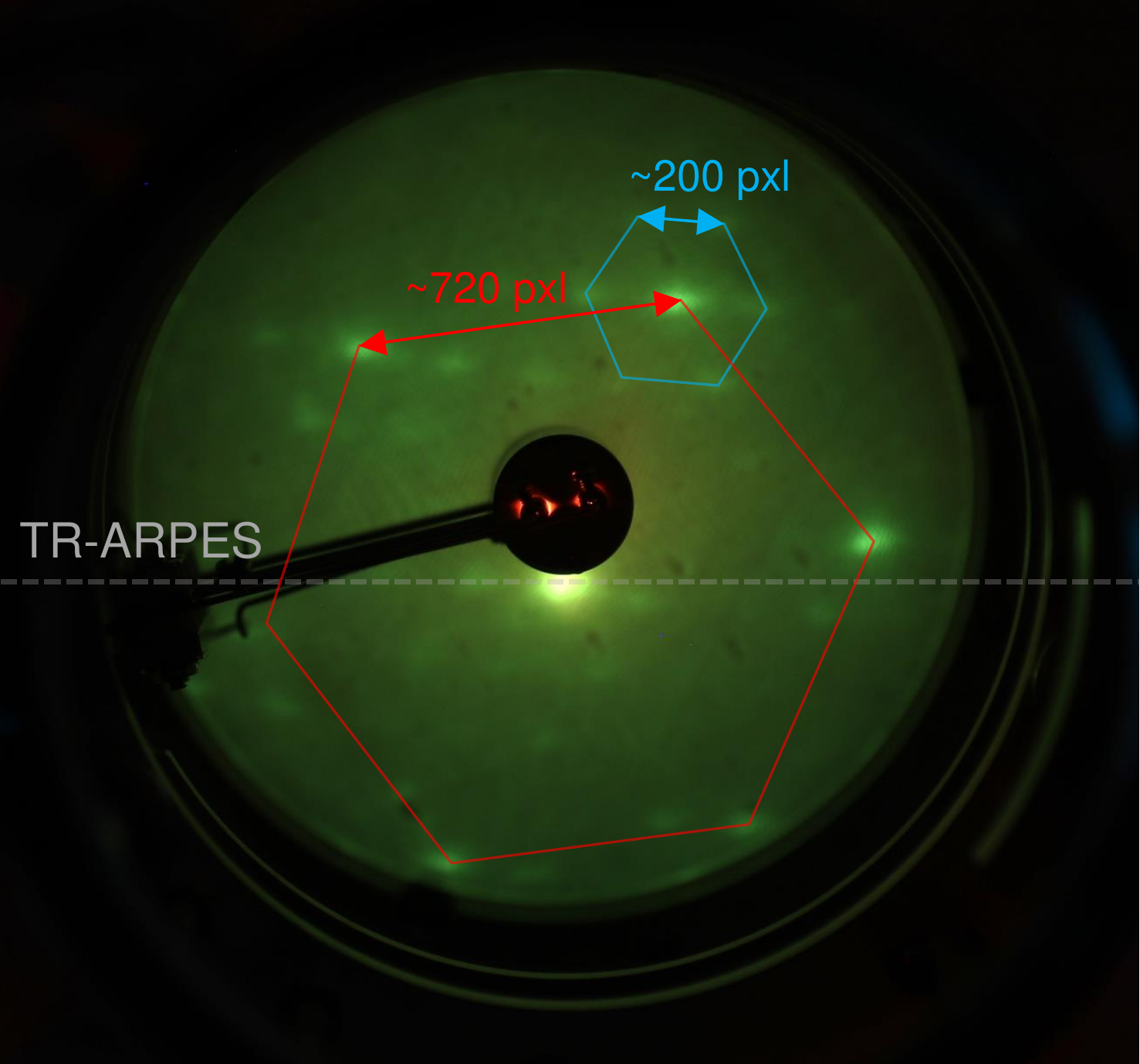}}
	{\caption{\label{fig:LEED} Low-energy electron diffraction (LEED). The pattern was obtained from a \TaSe~single crystal mounted on the sample holder in the analysis chamber prior to TR-ARPES measurements.}}
\end{figure}

Prior to TR-ARPES measurements, a low-energy electron diffraction (LEED) pattern of the \TaSe~sample was acquired in order to determine the orientation of the crystal as shown in Fig.~\ref{fig:LEED}. This allowed an estimation of the measured momentum direction, \kpara~which is aligned with the lab frame horizontal, as illustrated by the dashed line. Based on this, we obtained TR-ARPES spectra approximately along the \MBar-\GammaBar-\MBar~direction.

The hexagonal symmetry (red sketch) of \TaSe~is clearly visible from the diffraction spot pattern. On each of the corners of this main feature, weak secondary spots were observed (blue sketch). The ratio of the spot separation between these features gives 720 pxl / 200 pxl = 3.6 $\approx \sqrt{13}$. The secondary feature is also rotated by $\sim$ 13$^{\circ}$ with respect to the primary. Hence, the secondary spots come from the \PLDsqrtthir~superlattice structure with 13$^{\circ}$ rotation as expected in the CDW phase \cite{suppWilson1975,suppAebi2001,suppBovet2004}. The same features were found at room temperature and 77 K.

\section{Full-wavevector ARPES}\label{Sec:Full-wavevector ARPES}
\begin{figure}[h]
	\begin{center}
		\includegraphics[width=0.85\linewidth,clip]{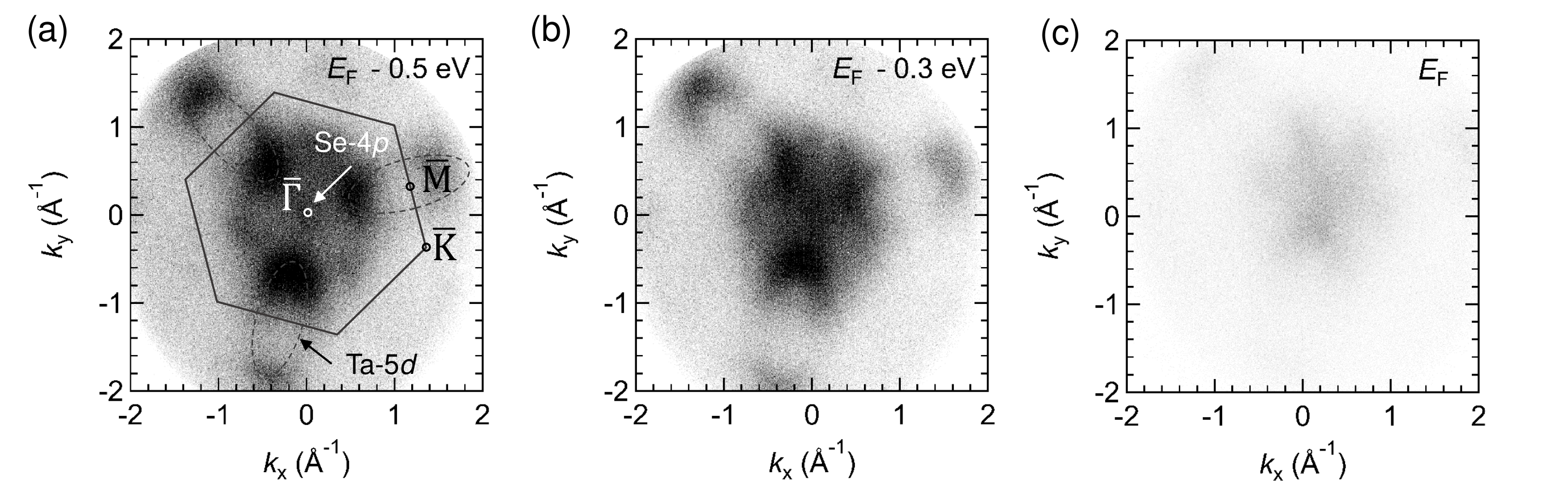}
	\end{center}
	\vspace*{-0.2in}
	\caption{\label{fig:ARPES} Full-wavevector ARPES (\hv~= 21.2 eV) images of \TaSe~at 40 K at specific binding energies as indicated. Panel (a) includes a sketch of the BZ (black hexagon) and an outline of the Ta-5\textit{d} derived electron pockets (grey dashed ellipse). The labelled high-symmetry points are a projection onto the experimental $k_{z}$.}
\end{figure}

Fig.~\ref{fig:ARPES} shows full-wavevector ARPES images at specific binding energies. At (\Ef~- 0.5) eV in Fig.~\ref{fig:ARPES}(a), the main features of the electronic structure of \TaSe~can be seen. On the edges of the BZ around the \MBar-point, are the elliptical Ta-5\textit{d} derived electron pockets. At 40 K, far below the CDW transition temperature (473 K), there is a loss of intensity on the parallel arms of these pockets where the CDW gap is excepted to occur \cite{suppWilson1975}. At the BZ centre (\GammaBar-point), a weak spot is visible which originates from the top of the Se-4\textit{p} derived pocket. In addition, there is a broad \textit{pancake} feature which extends across a large portion of the BZ from the centre to the inner corners of the electron pockets. At (\Ef~- 0.3) eV in Fig.~\ref{fig:ARPES}(a), the electronic structure is still clearly visible. However at the Fermi level, \Ef~in Fig.~\ref{fig:ARPES}(a), there is almost a complete loss of intensity across all $k$-space. This provides evidence that the entire Fermi surface is removed at 40 K as a result of the Mott transition.

\section{Fluence dependence of valence band dynamics}\label{Sec:Fluence dependence of valence band dynamics}
\begin{figure}[H]
	\floatbox[{\capbeside\thisfloatsetup{capbesideposition={right,center},capbesidewidth=5.5cm}}]{figure}[\FBwidth]
	{\includegraphics[width=10cm]{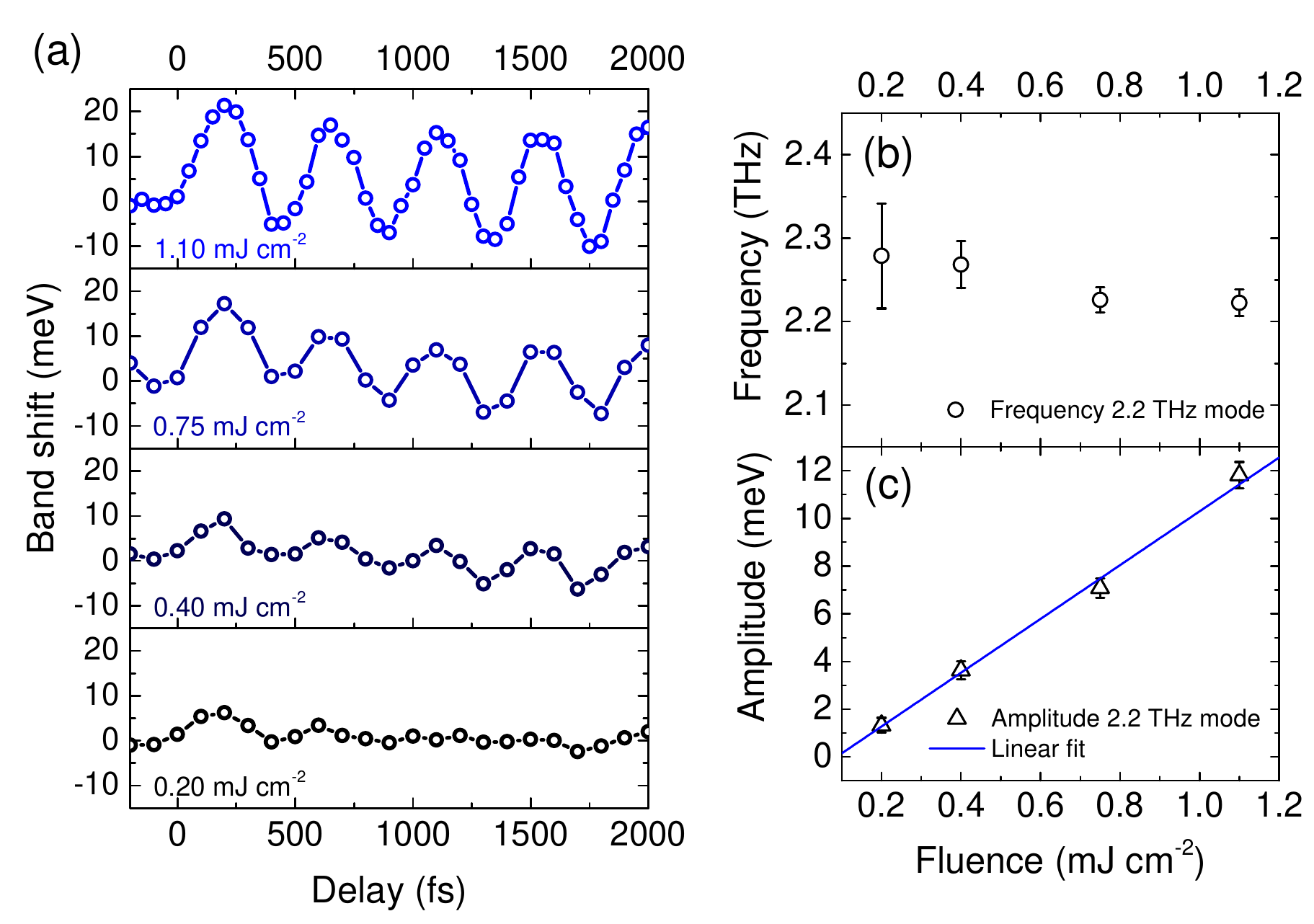}}
	{\caption{\label{fig:Fluence} Fluence dependence of valence band dynamics at 77 K. (a) Shift of the valence band edge. Frequency (b) and amplitude (c) of the oscillatory component of the valence band edge as a function of pump fluence.}}
\end{figure}

We performed TR-ARPES experiments at 77 K for a series of pump laser fluences in the range (0.20 - 1.10)~\mJcm. Fig.~\ref{fig:Fluence}(a) shows the dynamics of the valence band edge shift for all fluences, where clear oscillations are observed. Figs.~\ref{fig:Fluence}(b) and (c) show the frequency and amplitude of these oscillations respectively, which were obtained by fitting the data with a periodic function. Here, we did not consider the damping time, $\tau_d$ since it is larger than the temporal window of 2 ps ($\tau_d  \approx$ 6.3 ps, see main text). We find the same single-frequency oscillations ($\sim$2.2 THz) of the CDW amplitude mode as discussed in the main text, with no significant variation as a function of fluence. In addition, we find that the oscillation amplitude increases linearly with increasing fluence, thus confirming that the experiment was performed in a linear excitation regime i.e. avoiding multi-phonon processes (quadratic) or saturating behaviour induced by the pump.

\bibliographystyle{unsrtnat}

\end{document}